# Inner Cell Mass and Trophectoderm Segmentation in Human Blastocyst Images using Deep Neural Network

Md Yousuf Harun, *Student Member, IEEE*, Thomas Huang, and Aaron T. Ohta, *Senior Member, IEEE*

*Abstract*— Embryo quality assessment based on morphological attributes is important for achieving higher pregnancy rates from in vitro fertilization (IVF). The accurate segmentation of the embryo's inner cell mass (ICM) and trophectoderm epithelium (TE) is important, as these parameters can help to predict the embryo viability and live birth potential. However, segmentation of the ICM and TE is difficult due to variations in their shape and similarities in their textures, both with each other and with their surroundings. To tackle this problem, a deep neural network (DNN) based segmentation approach was implemented. The DNN can identify the ICM region with 99.1% accuracy, 94.9% precision, 93.8% recall, a 94.3% Dice Coefficient, and a 89.3% Jaccard Index. It can extract the TE region with 98.3% accuracy, 91.8% precision, 93.2% recall, a 92.5% Dice Coefficient, and a 85.3% Jaccard Index.

## I. Introduction

In vitro fertilization (IVF) is an effective treatment for infertility. During IVF, embryos with the highest potential for a successful pregnancy are selected for transfer to mother's uterus. Traditional embryo selection is done by visual assessment of embryo morphology, which is subjective and time consuming. Therefore, there is a strong need for an automatic, objective method of embryo evaluation, such as an image analysis tool that can identify morphological features of the embryos with at least 32 cells (blastocysts) for IVF. According to Gardner blastocyst grading [1], the degree of blastocyst expansion, the characteristics of the inner cell mass (ICM), and the quality of the trophectoderm epithelium (TE) are the most important factors in predicting the pregnancy outcome. ICM is believed to be an effective parameter for embryo viability assessment because this eventually develops into a fetus [2]. TE is also important because successful hatching of an implanted embryo correlates highly with strong TE layer [3]. Hence, the quality of both the ICM and TE needs to be analyzed to evaluate embryonic potential. The shape of the ICM and TE are very irregular, and they both have similar textures. They are surrounded by two other irregularly shaped regions called the zona pellucida (ZP) and the cavity mass (CM) (Fig. 1). Thus, their shape variability and their similar textures make it challenging to segment the ICM and TE.

In this paper, we present a segmentation method based on a deep neural network (DNN) to detect the ICM and TE regions. We trained and optimized a U-Net-based DNN to extract both local (texture of the ICM and TE) and contextual information (spatial arrangement of the ICM and TE) to segment the ICM and TE regions. The results from the DNN are compared with current state-of-the-art methods to demonstrate its efficacy.

## II. Related Works

There have been many attempts to automate embryo segmentation [4], [5], [6], including works on the automatic segmentation of the ICM and TE. Santos et al. [7] demonstrated a semi-automatic segmentation method for the ICM and TE using level-sets. Since the performance of level-sets relies on predefined initial contour, this method does not work if the ICM resides outside of the contour. The viability of this approach was not confirmed with quantitative results. Singh et al. [8] showed an automatic segmentation method for the TE region by utilizing the Retinex and level-set algorithms, resulting in 87.8% accuracy and 78.7% recall. Kheradmand et al. [9] introduced a two-layer feedforward backpropagation neural network to segment the ZP, TE, and ICM by exploiting the discrete cosine transform (DCT). In this method, the ICM, TE, and ZP segmentations are limited to Jaccard Indices of 47.7%, 58.9%, and 67.4%, respectively. Moradi et al. [10] presented an automatic ICM identification method based on coarse-to-fine texture analysis. The ICM is first detected using Gabor and DCT features, and then a region-based level-set algorithm is utilized to finalize the ICM boundaries. This method achieved a Jaccard Index of 70.3%. Saeedi et al. [11] presented a texture-based method for the automatic segmentation of the ICM and TE. This method utilizes various texture information along with k-means clustering and watershed segmentation algorithms. This method achieved a 71.1% Jaccard Index for ICM segmentation and a 63% Jaccard Index for TE segmentation. Kheradmand et al. [12] presented a fully convolutional network based method, where they used a pretrained 16-layer visual geometry group network [13] for ICM segmentation. The best Jaccard Index achieved by this method was 76.5%. Recently, Moradi et al. [14] introduced a novel U-Net variant and a multi-resolutional method for ICM segmentation. Utilizing dilated convolution [15], a stack of five dilated convolution layers is incorporated into the central bridge part of U-Net. This increased the network's receptive field by 40% and outperformed the previous methods [9-12]. However, the reported precision, recall and Dice Coefficient were still less than 92%, and the Jaccard Index was below 82%. Since the ICM and TE are crucial factors in assessing the embryo quality, a more robust method with increased performance is needed.

Md Yousuf Harun and Aaron T. Ohta are with the Electrical Engineering Department, University of Hawaii at Manoa, Honolulu, HI, USA (e-mail: mdyousuf@hawaii.edu, aohta@hawaii.edu).

Thomas Huang is with the Pacific IVF Institute, Kapiolani Medical Center, Honolulu, HI, USA (e-mail: huangt@hawaii.edu).

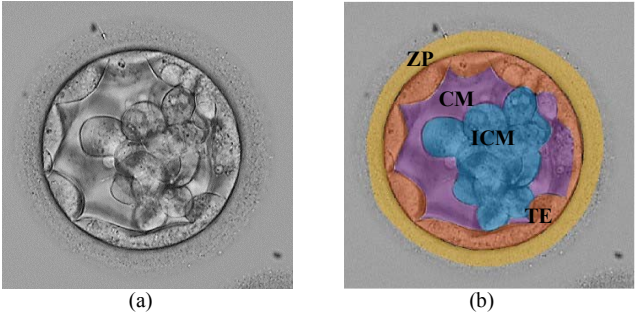

Figure 1. Images of (a) a blastocyst and (b) its annotated components.

III. METHODOLOGY

*A. Method Overview*

A supervised learning approach was used to train a DNN using labeled samples (ground truths). The network learns the invariant and discriminatory features of the ICM and TE from the training set and segments them in the test set. The segmentation performance is enhanced by optimizing the network, fine tuning hyperparameters, and preprocessing and augmenting the dataset.

*B. Network Structure*

U-Net [16] and its variants have become a popular choice for biomedical image analysis since they can provide decent performance with relatively small dataset. U-Net is composed of a contracting path (encoder) and an expansive path (decoder), connected via skip connections. The encoder downsamples the input image into a representative feature maps and the decoder upsamples the input feature maps into a pixel-wise categorization. Inspired by Residual U-Net [17], Dilated Residual U-Net [18] and Residual Dilated U-Net [19], we implemented a similar model to Ref. [19] for this task. This includes the advantages of U-Net skip connections [20], residual learning [21], and dilated convolution [15]. We customized the network to be effective on the blastocyst dataset. The depth of encoder-decoder is 4. Each encoder unit consists of one convolution block, one subsampling block, and one residual block. The first encoder through fourth encoder have 16, 32, 64, and 128 kernels, respectively. The bottleneck consists of total 4 dilated convolution layers with dilation rates of 1, 2, 4, and 8, as illustrated in Fig. 2. Each decoder unit consists of one up-convolution block followed by a concatenation with the encoder, one convolution block, and one residual block. The first decoder through fourth decoder have 128, 64, 32, and 16 kernels, respectively. A batch normalization layer was added after each upsampling layer in the decoder. This improves the network's performance [22].

IV. EXPERIMENTS

*A. Dataset*

We used the publicly available blastocyst dataset introduced by Saeedi et al. [11]. The ICM and TE regions in the images were manually annotated by experts at Pacific Centre for Reproductive Medicine (PCRM), Canada. The annotated images are used as ground truth to train and test the DNN. This dataset has 249 images in total. We split the total dataset into a training set (85% of the images) and a test set (15% of the images).

*B. Experimental Setup*

The DNN is trained and tested using an NVIDIA GeForce GTX 1070 GPU with 8 GB of memory and 16 GB of RAM. The model is implemented using Keras with a Tensorflow backend. We used a minibatch size of 16 and maximum epochs of 200.

*C. Data Preprocessing, Augmentation, and Randomization*

Since the images in the dataset have different resolutions, we resized them to a uniform 256 × 256 resolution. Then, we applied standard normalization. Given the complexities involved in the segmentation task, particularly the similar textures and irregularities in the shapes, the relatively small training set was augmented to improve the DNN's generalization. Here, we adopted the dataset augmentation technique from Ref. [12]. Each image in training set was rotated incrementally by 10 degrees, up to a full 360-degree rotation; thus, a single image becomes 36 transformed images. Due to the elliptical nature of the blastocyst morphology, the original data distribution was unaffected by this process. To prevent the network from memorizing the dataset, the training and test sets were randomized. The training set images were also randomized during each training iteration.

*D. Hyperparameter Optimization*

We utilized binary cross entropy Jaccard Index loss to compensate for class imbalance (ICM and TE, and non-ICM and non-TE). The loss function is minimized by the ADAM optimizer. The initial learning rate was set to 0.0001. We added a 5% learning rate reduction with a patience of 5 epochs using the Keras callback function. This helped to overcome overshooting the minimum, and improved convergence. Furthermore, we included the early stopping callback with a patience of 15 epochs to prevent overfitting. The callback monitored the loss function to optimize the training process. A 5% dropout was also added for a similar purpose. The Jaccard Index did not vary significantly for threshold values from 0.4 to 0.6, so a 0.5 threshold was used for the final prediction.

V. EVALUATION AND RESULTS

*A. Evaluation Criteria*

To evaluate the network's performance, the ICM and TE segmentations were compared with their ground truth annotations. We used five evaluation criteria: accuracy, precision, recall, Dice Coefficient, and Jaccard Index. These metrics [23-26] are defined based on four parameters: TP (True Positive), FP (False Positive), TN (True Negative), and FN (False Negative). Here, TP measures the number of pixels that are correctly identified as the ICM or TE region. TN shows the number of pixels that are truly detected as background (non-ICM or non-TE) pixels. FP indicates the number of pixels that are incorrectly classified as the ICM or TE region. FN counts the misclassified background pixels.

Accuracy represents the proportion of pixels correctly extracted as the background or the ICM or TE:

$$Accuracy = \frac{TP + TN}{TP + TN + FP + FN} \qquad (1).$$

Precision is the fraction of predicted ICM or TE pixels which are labeled as the ICM or TE regions:

$$Precision = \frac{TP}{TP + FP} \quad (2).$$

Recall is the fraction of all the labeled ICM or TE pixels that are correctly predicted:

$$Recall = \frac{TP}{TP + FN} \quad (3).$$

The Dice Coefficient denotes the similarity between the automated and ground truth segmentations:

$$Dice\ Coefficient = \frac{2 \times TP}{2 \times TP + FP + FN} \quad (4).$$

The Jaccard Index measures the true pixels retrieved only for the ICM or TE regions:

$$Jaccard\ Index = \frac{TP}{TP + FP + FN} \quad (5).$$

The Jaccard Index and the Dice Coefficient are better at dealing with the class imbalance. Both metrics range from 0 to 1 with 1 signifying perfectly overlapping segmentation.

B. *Comparison with the State-of-the-Art*

The results from the DNN are compared with existing state-of-the-art methods in Tables 1 and 2. The experimental results confirm that the DNN achieves higher performance than previously reported results [8-12, 14]. ICM segmentation results show that our method outperforms the results from Ref. [14] by 0.8% in accuracy, 7.1% in precision, 2.5% in recall, 5.4% in the Dice Coefficient, and 9.4% in the Jaccard Index. TE segmentation results indicate that the DNN method outperforms Ref. [11] by 13.5% in accuracy, 33% in precision, 4.7% in recall, 19.7% in the Dice Coefficient, and 35.4% in the Jaccard Index.

C. *Segmentation Results Verification*

To verify the segmentation results, the predicted ICM and TE are compared with the manually labeled ICM and TE. The extracted ICM and TE ground truth boundaries are overlaid on the segmented ICM and TE to visualize the difference. To better understand the quality of the ICM segmentation, we categorize the results according to best (Jaccard Index of more than 97%), better (Jaccard Index from 92% to 97%), and fair (Jaccard Index from 77% to 92%) segmentation, as shown in Table 3. Of the test set images, 36.8% are in the best segmentation category, 50% are in the better segmentation category, and the remaining 13.2% are in the fair segmentation category. Table 4 shows the TE segmentation results according to best (Jaccard Index of more than 94%), better (Jaccard Index from 87% to 94%), and fair (Jaccard Index from 76% to 87%) segmentation categories. Of the test set images, 31.6% are in the best segmentation category, 47.4% are in the better segmentation category, and the remaining 21% are in the fair segmentation category. The segmentation results have been ranked by the Jaccard Index and the Dice Coefficient since other performance metrics are reasonably high.

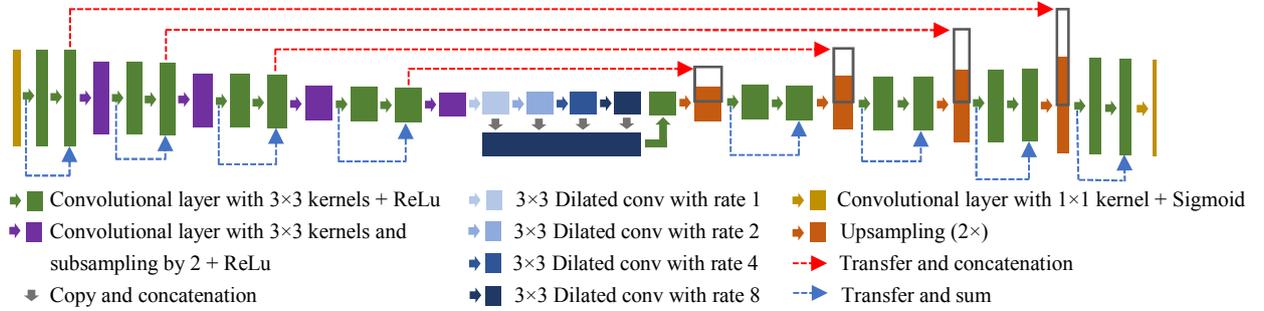

- ➡ ▮ Convolutional layer with 3×3 kernels + ReLu
- ➡ ▮ Convolutional layer with 3×3 kernels and subsampling by 2 + ReLu
- ➡ Copy and concatenation
- ➡ ▮ 3×3 Dilated conv with rate 1
- ➡ ▮ 3×3 Dilated conv with rate 2
- ➡ ▮ 3×3 Dilated conv with rate 4
- ➡ ▮ 3×3 Dilated conv with rate 8
- ➡ ▮ Convolutional layer with 1×1 kernel + Sigmoid
- ➡ ▮ Upsampling (2×)
- ⇢ Transfer and concatenation
- ⇢ Transfer and sum

Figure 2. DNN structure.

**Table 1.** Comparison of ICM results of this paper with state-of-the-art methods based on same dataset.

| Method | Accuracy (%) | Precision (%) | Recall (%) | Dice Coefficient (%) | Jaccard Index (%) |
|---|---|---|---|---|---|
| Kheradmand et al. [9] | 93.0 | 75.6 | 56.4 | 64.6 | 47.7 |
| Moradi et al. [10] | -- | 78.7 | 86.8 | 82.6 | 70.3 |
| Saeedi et al. [11] | 93.3 | 84.5 | 78.3 | 83.1 | 71.1 |
| Kheradmand et al. [12] | 95.6 | -- | -- | 86.7 | 76.5 |
| Moradi et al. [14] | 98.3 | 88.6 | 91.5 | 89.5 | 81.6 |
| This paper | **99.1** | **94.9** | **93.8** | **94.3** | **89.3** |

**Table 2.** Comparison of TE results of this paper with state-of-the-art methods based on same dataset.

| Method | Accuracy (%) | Precision (%) | Recall (%) | Dice Coefficient (%) | Jaccard Index (%) |
|---|---|---|---|---|---|
| Singh et al. [8] | 86.7 | 71.3 | 83.1 | 76.7 | 62.2 |
| Kheradmand et al. [9] | 90.0 | 69.1 | 80.0 | 74.2 | 58.9 |
| Saeedi et al. [11] | 86.6 | 69.0 | 89.0 | 77.3 | 63.0 |
| This paper | **98.3** | **91.8** | **93.2** | **92.5** | **85.3** |

**Table 3.** ICM segmentation results. The background (non-ICM region) is colored dark cyan, the annotated ICM (ground truth) is light green, the DNN-segmented ICM is yellow, and the contour of the ground truth is red. JI and DC stand for Jaccard Index and Dice Coefficient, respectively.

| | Original Image | Ground Truth | Segmentation | Verification | |
|---|---|---|---|---|---|
| Best Segmentation (Jaccard > 97%) | 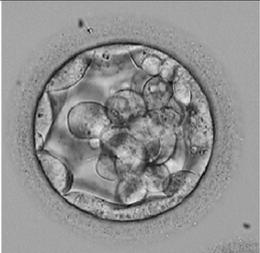 | 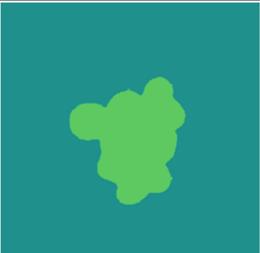 | 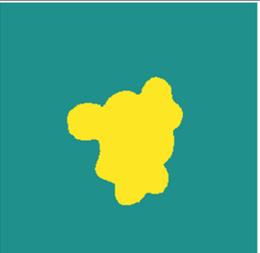 | 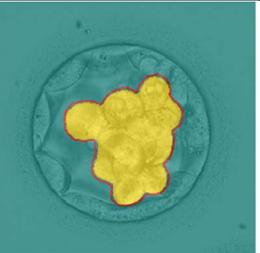 | JI 97.9% (highest), DC 99.1% |
| | 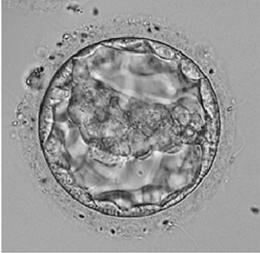 | 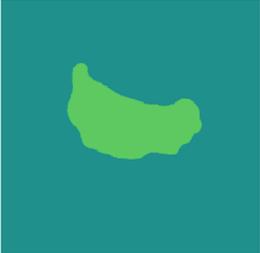 | 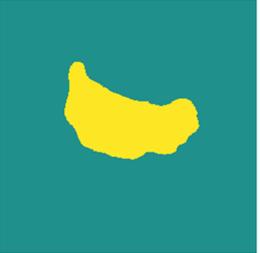 | 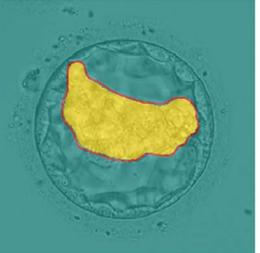 | JI 97.6%, DC 98.8% |
| Better Segmentation (Jaccard > 92%) | 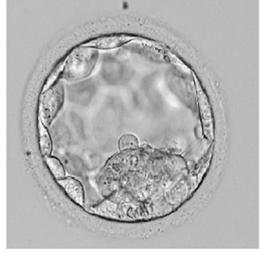 | 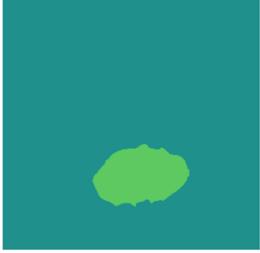 | 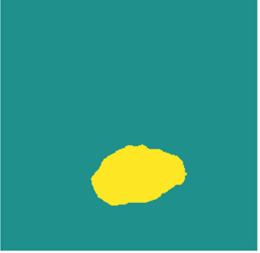 | 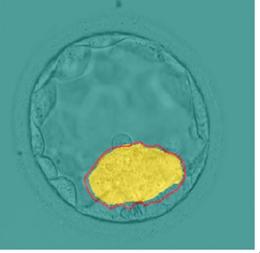 | JI 92.9%, DC 96.3% |
| | 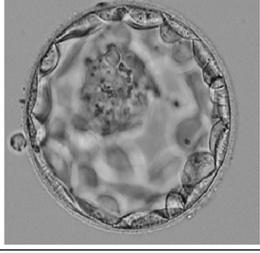 | 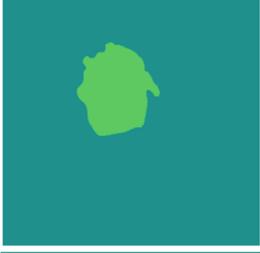 | 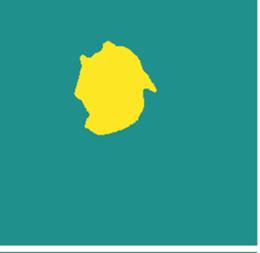 | 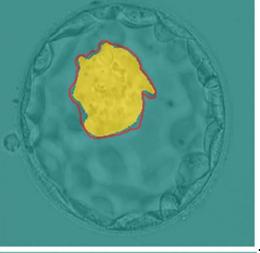 | JI 94%, DC 97.1% |
| Fair Segmentation (Jaccard > 77%) | 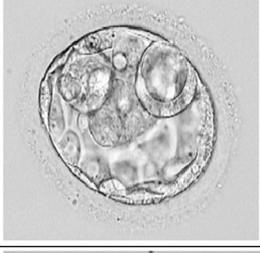 | 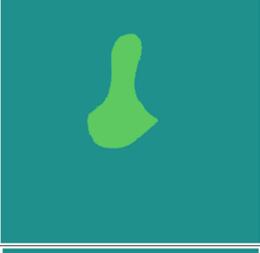 | 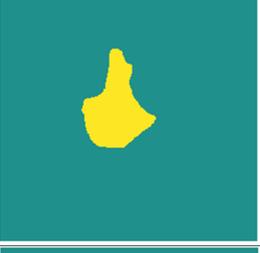 | 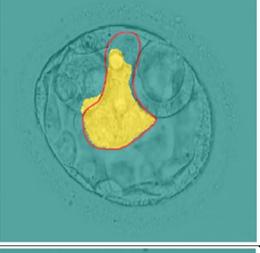 | JI 77.4% (lowest), DC 87.3% |
| | 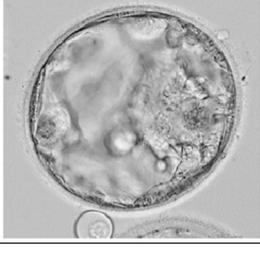 | 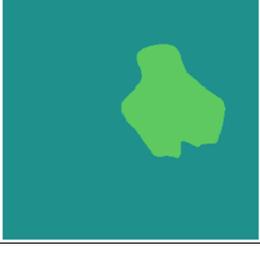 | 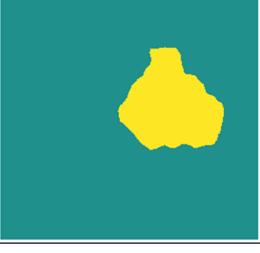 | 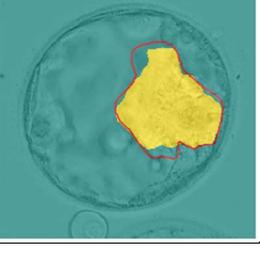 | JI 86.2%, DC 92.6% |

**Table 4.** TE segmentation results. The background (non-TE region) is colored dark cyan, the annotated TE (ground truth) is light green, the DNN-segmented TE is yellow, and the contour of the ground truth is red. JI and DC stand for Jaccard Index and Dice Coefficient, respectively.

| | Original Image | Ground Truth | Segmentation | Verification | |
|---|---|---|---|---|---|
| Best Segmentation (Jaccard > 94%) | | | | | JI 95.2% (highest), DC 97.9% |
| | | | | | JI 94.9%, DC 97.8% |
| Better Segmentation (Jaccard > 87%) | | | | | JI 87.4%, DC 93.6% |
| | | | | | JI 89.2%, DC 95.5% |
| Fair Segmentation (Jaccard > 76%) | | | | | JI 76.7% (lowest), DC 87.1% |
| | | | | | JI 78%, DC 88.2% |

## VI. Conclusion

We presented a segmentation method using a deep learning algorithm to precisely identify two vital components of blastocysts: the inner cell mass and the trophectoderm. This DNN method achieved a 94.3% Dice Coefficient and a 89.3% Jaccard Index for ICM segmentation, and a 92.5% Dice Coefficient and a 85.3% Jaccard Index for TE segmentation. This demonstrates the robustness and reliability of the DNN method. This work can be used for more accurate identification of blastocyst components and for automated quality assessment of human embryos.

## Acknowledgment

We thank Prof. Parvaneh Saeedi from Simon Fraser University, Canada, for providing access to the dataset used in this paper.